\documentclass[preprint,12pt]{elsarticle}

\usepackage{graphicx}
\usepackage{amssymb}
\usepackage{booktabs}
\usepackage{multirow}
\usepackage{algorithm}
\usepackage{algpseudocode} 
\usepackage{amsmath}
\algnewcommand\algorithmicforeach{\textbf{for each}}
\algdef{S}[FOR]{ForEach}[1]{\algorithmicforeach\ #1\ \algorithmicdo}
\usepackage{url}
\usepackage{float}

\begin{document}

\begin{frontmatter}

\title{Structural balance in signed digraphs: considering transitivity to measure balance in graphs constructed by using different link signing methods}

\author{Ly Dinh*, Rezvaneh Rezapour*, Lan Jiang, Jana Diesner}
\address{University of Illinois at Urbana--Champaign}
\address{* These authors contributed equally to this work}
\ead{\{dinh4,rezapou2,lanj3,jdiesner\}@illinois.edu}
\begin{abstract} 

Structural balance theory assumes triads in networks to gravitate towards stable configurations. The theory has been verified for undirected graphs. Since real-world networks are often directed, we introduce a novel method for considering both transitivity and sign consistency for calculating balance in signed digraphs. We test our approach on graphs that we constructed by using different methods for identifying edge signs: natural language processing to infer signs from underlying text data, and self-reported survey data. Our results show that for various social contexts and edge sign detection methods, balance is moderately high, ranging from 67.5\% to 92.4\%. 

\end{abstract}


\end{frontmatter}

\section{Introduction}
\label{S:1}

Real-world social and communication networks are composed of complex and continually evolving interactions among social agents. Network scholars have examined core principles that explains patterns of social interactions at various levels of analysis, i.e., at the node level \cite{borgatti2006graph}, dyadic level \cite{block2015reciprocity}, triadic level \cite{cartwright1956structural, johnsen1986structure}, subgroup level \cite{riley1986social}, and the graph level \cite{kilduff2003social}. Each level of analysis enables different types of hypotheses to be tested, e.g., with respect to structural properties, such as reciprocity at the dyad level, transitivity at the triad level, clusterability at the subgroup level, and centralization at the network level \cite{wasserman1994social,monge2003theories}. 
In this study, we focus on a fundamental unit of analysis, namely triads, given that there are patterns of tie formations between three actors and their associated ties that cannot be explained at any other levels of analysis \cite{simmel1950stranger, flament1963applications, taylor1970balance}. Some of the basic dynamics and theories that drive and explain interactions at the triadic level are structural balance \cite{heider1946attitudes} and transitivity  \cite{feld1982patterns}, which both model
how stable relationships emerge in groups of three nodes. Furthermore, extant literature on structural balance has considered signed and undirected triads for analysis \cite{harary1980simple,newcomb1968interpersonal,krackhardt2006heider}; an approach that models relationships between any three individuals as being reciprocated. In real-world networks, however, individual $P$ may perceive individual $O$ as a friend, but $O$ may not have the same perception of $P$. This simple example highlights the importance of considering directionality for understanding balance in signed networks. Revisiting the central tenet of structural balance theory \cite{heider1946attitudes, doreian2001pre}, we propose that directionality can be understood in terms of the transitive properties within a triad. Therefore, this paper is based on the assumption that a signed and directed triad is balanced if every transitive semicycle within the triad is positive.

We propose that transitivity is a necessary pre-condition to evaluate balance in signed digraphs. Stemming from the Heiderian \cite{heider1946attitudes, heider1958psychological} assumption transitivity is a crucial property that explains how directed and signed ties are oriented in ways that are consistent with balance. Therefore, we incorporate both conditions of transitivity and sign consistency to the assessment of structural balance for triads in a signed digraph. In particular, we leverage the triad census \cite{holland1978omnibus} to extract all transitive semicycles within a particular triad, and calculate overall balance with respect to the proportions of balanced semicycles within the triad. We test this method of calculating structural balance on three different signed digraphs with three different edge types, namely sentiment \cite{diesner2015little, hallinan1974structural}, morality, and perceived trust \cite{van2005evolution}. Analyzing networks with different edge types enables us to determine if balance is consistent within and across different types of relations.

Our analysis shows that balance ratios vary across different measurements of social relations, with the average balance ratio based on morality being 81.7\%, based on sentiment being 69.5\%, and based on perceived trust being 72.7\%. One commonality across the networks is that balance ratios are high (70\% and above), which offers an empirical validation of balance theory. 

This paper makes three contributions: First, we extend the theory of structural balance to include signed digraphs where both transitivity and sign consistency are required and considered for calculating balance in triads with signed and directed edges. This helps to model communication networks and other social networks where ties might be directed in a more comprehensive way. Second, we apply two different methods for identifying edge signs: natural language processing to infer two different types of edge signs from data authored by nodes, and surveys to elicit self-reported data from nodes about edge signs. Third, we empirically assess balance in two different and contemporary contexts, namely remote communications in two business organizations, and team-based interactions in a virtual environment. 

The paper proceeds as follows: the first three sections discuss previous literature and theoretical development towards an operationalization of balance for signed and directed networks that takes transitivity into account. In the next two sections, we discuss the methodology and empirical results for our balance analysis. The discussion and conclusion reflect on the contribution of our methodology to extend the analysis of balance for signed and directed networks.

\section{Related work}
The central tenet of social network analysis is to understand the structures of relations between sets of objects, such as individuals, groups, or organizations \cite{wasserman1994social}. Heider's structural balance theory \cite{heider1946attitudes} was one of the earliest formulations of how relationships form between three individuals, or between a pair of indivduals and their perceptions of or attitude towards an common object. Heider initially examined $POX$ triads, where $P$ is a focal individual, $O$ is a second individual, and $X$ is either an individual or a common object. He asserted that the relationship between these three entities is `balanced' if $P$ liked $O$, $O$ liked $X$, and $P$ also liked $X$. On the other hand, `imbalance' would occur if $P$ liked $O$, $O$ liked $X$, but $P$ did not like $X$. Heider's primary claim was that balance is a state of equilibrium, and that individuals in networks strive to move towards and maintaining that equilibrium or balance. In a later study, Heider proposed that balance co-exists with symmetry of relations \cite{heider1958psychological}, such that $P$ liking $O$ also implies $O$ liking $P$. 
Cartwright and Harary (1956) brought balance theory into the context of signed networks, where relationships between pairs of nodes can be represented as either a ($+$) or ($-$). They further formalized that a triad within a network is balanced if the product of the signs of its edges is positive. The four possible types of triads in the context of balance assessment are shown in table \ref{tab:Table1}, showing the structure of balance and imbalance.

\begin{table}[H]
\centering
\begin{tabular}{@{}cccc@{}}
\toprule
P—O & O—X & P—X & Characteristic \\ \midrule
+ & + & + & Balanced \\
+ & + & - & Imbalanced \\
+ & - & + & Imbalanced \\
+ & - & - & Balanced \\
- & + & + & Imbalanced \\
- & + & - & Balanced \\
- & - & + & Balanced \\
- & - & - & Imbalanced \\ \bottomrule
\end{tabular}
\caption[position=bottom]{Patterns of signed relations in $POX$ triad based on balance theory}
\label{tab:Table1}
\end{table}

In addition to symmetry as a prerequisite for balance, in the $POX$ triad as stated above, Heider posited that ``three positive relations may be considered psychologically transitive'', in that ``$P$ tends to like $X$ if $P\mathcal{R}O$ and $O\mathcal{R}X$ are valid at the same time'' ($\mathcal{R}$ represents positive relation between two nodes) \cite{heider1946attitudes}. Transitivity is thus established as a necessary condition for stability \cite{hallinan1975higher} and balance \cite{holland1978omnibus, krackhardt2006heider} in a social network. Davis, Holland, and Leinhardt \cite{davis1979davis} in their studies of positive relations in triads found that transitivity is a pervasive property in balanced (all positive) triads. Stix \cite{stix1974improved} also asserted that transitivity is an important property of balance, and that ``the transitivity of any structure corresponds to its sensitivity to imbalance'' (p. 447). In addition to the condition of symmetry, transitivity plays a vital role in explaining the formation of ties within triads. 

\begin{figure}[htb]
\centering
\centerline{\includegraphics[width=\linewidth]{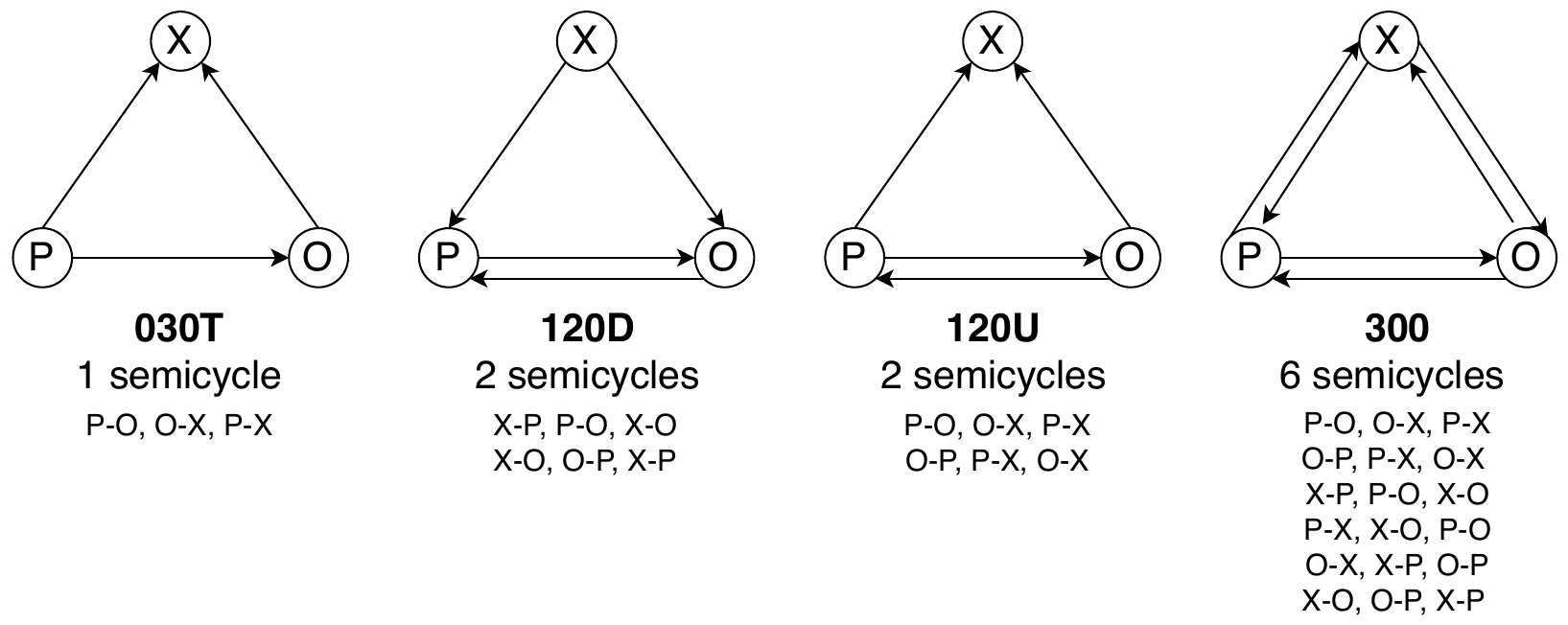}}
\caption{Triad census (include only triads which are transitive and balanced)}
\label{fig:Fig1}
\end{figure}

\begin{figure}[t]
\centering
\centerline{\includegraphics[width=\linewidth]{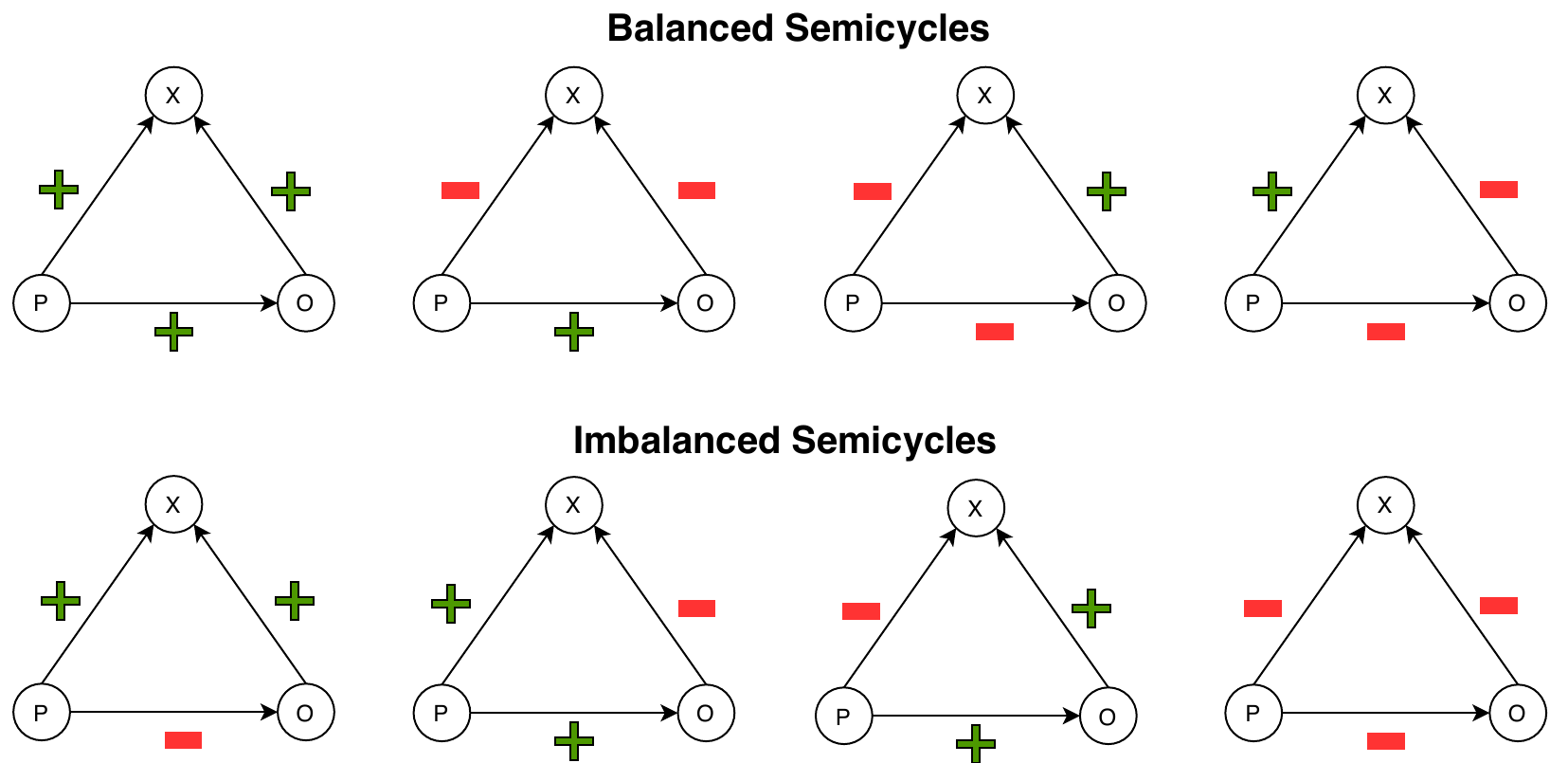}}
\caption{Balanced and imbalanced semicycles}
\label{fig:Fig2}
\end{figure}

In real-world networks, however, relations might not be reciprocated \cite{wasserman1994social}. For instance, $P$ may regard person $O$ as a friend, but $O$ may not see $P$ as a friend. In such cases, ties are more appropriately represented as directed than undirected edges. In response, scholars have refined the assessment of balance to the level of the semicycles (containing directed ties) embedded each triad \citet{feather1964structural,roberts1974structural,de1999sign}, and the extent to which each semicycle is balanced. A semicycle is balanced if the product of its edges is positive (examples of balanced and imbalanced semicycles are shown in figure \ref{fig:Fig2}). Furthermore, semicycles that are cyclic ($P$ $\rightarrow$ $O$, $O$ $\rightarrow$ $X$, $X$ $\rightarrow$ $P$) are not suitable for balance analysis because they are intransitive \cite{roberts1974structural,block2015reciprocity}. Another reason to not consider cyclic configuration for balance assessment is because cycles contained limited information on the process of influence among relationships \cite{rank2010structural,veenstra2013network}. Another effort to integrate directed edges into triadic analysis stems from Holland and Leinhardt \cite{holland1978omnibus}, who developed sixteen classes of MAN (Mutual, Asymmetric, Null) triads, also known as the triad census. These classes represent all possible combination of directed ties between three nodes. Specifically, each configuration contains different combinations of edges, either it be mutual ($P$ likes $O$ and $O$ likes $P$), asymmetric ($P$ likes $O$ but $O$ does not like $P$), or null ($P$ and $O$ do not like each other). The triad census characterizes four triad types that are driven by both transitivity and balance (as show in figure \ref{fig:Fig1}). As we will show, these four configurations are relevant for our operationalization of balanced triads with transitivity as a precondition of balance. 

While a number of theorems has been developed to incorporate directionality into the calculation of balance, studies have ignored the direction of ties and thus analyzed balance for undirected networks as opposed to directed ones. In this study, we revisit the literature on semicycle balance to develop and demonstrate a solution for calculating structural balance for signed and directed networks while considering both transitivity and edge signs. 

\section{Expansion of balance theory using transitivity and direction of edges} \label{RQ}

In this section, we present our approach to calculating balance in signed directed networks with respect to transitive semicycles.

\section{Problem definition and notations} \label{theory}

Let $G$ be a signed digraph where $G = (D,\sigma)$. $D$ is a digraph underlying $G$, where $\mathcal{D} = (V,E)$ and sign function $\sigma : E \rightarrow \{+,-\}$. A \textit{triad} $T$ in $G$ is a set of three nodes with one directed edge between each two of them (in either direction). 

Definition 1: A \textit{semicycle} $S$ in signed directed $T$ is a set of three directed edges that starts from a vertex $V$, follows the direction of edges, and does not return to the same vertex. In other words, $S$ is \textit{transitive} and \textit{non-cyclic}. 

Definition 2: Every \textit{semicycle} $S$ in a signed digraph $G$ must be \textit{transitive} in order to be considered for our balance analysis. 
For transitivity, we consider four types of triads, i.e., (030T, 120D, 120U, and 300), because they contain only transitive semicycles (030T contains 1 transitive semicycle, 120D and 120U each contains 2 transitive semicycles, 300 contains 6 transitive semicycles). For simplicity of notation, we define $T^{(i)}$ as the set of all transitive triads of type $i$, where $i = 1, 2, 3, 4$ refers to 030T, 120D, 120U, and 300, respectively. 

As an example, consider triad type 300 between ($P,O,X$), where there are six permutations of $P,O,X$ that are transitive, from the point of view of each node. This also applies to triad types 120D (two permutations), 120U (two permutations), and 030T (1 permutation). 

\underline{Proposition:} We define $T$ a \textit{completely balanced} triad if and only if every transitive semicycle in $T$ is balanced (positive). A transitive semicycle is positive if it contains an even number of negative directed edges. Furthermore, we define $T$ a \textit{partially balanced} triad if it contains at least one negative semicyle. Finally, $T$ is \textit{completely imbalanced} if every transitive semicycle in $T$ is imbalanced (negative). 

Algorithm \ref{meth:algorithm} shows our step-by-step computation of balance in a signed directed network. 
After calculating balance $B_{T^{(i)}}^{j}$ for each triad $j$ of type $i$, we compute the balance ratio for the set of all transitive triads of type $i$ ($B_{T^{(i)}}$). Finally, the \textit{overall balance ratio of $G$} ($B_{Avg(G)}$) is calculated by averaging the balance ratio of all types $i$ across a network.  
A signed digraph $G = (D,\sigma)$ is balanced if all triads $T$ in $G$ are balanced. 

\begin{algorithm}[tbh]
\begin{algorithmic}[1]
\For{$ i = 1, 2, 3, 4 $}
\State Consider set $T^{(i)}$
\State $\rhd$ Take element $j$ of $T^{(i)}$, for $j = 1,\cdots, N_j$:
\State $\quad$ Find the Semicycles and calculate: $B_k^{sign} := \prod_r$ sign of edge $r$
\State $\quad$ Consider $S_{T^{(i)}}^{+,j} =$ \{semicycle $k$, $B_k^{sign}$ is $+$\}
\State  $\quad$ Consider $S_{T^{(i)}}^{-,j} =$ \{semicycle $k$, $B_k^{sign}$ is $-$\}
\State $\quad$ Let $S_{T^{(i)}}^{j} = S_{T^{(i)}}^{+,j} \cup S_{T^{(i)}}^{-,j}$
\State $\rhd$ Define:
\begin{equation*} B_{T^{(i)}}^{j} := \frac{|S_{T^{(i)}}^{+,j}|}{|S_{T^{(i)}}^{-,j}|}, \quad \left(\text{Note: } B_{T^{(i)}}^{j} \in [0,1]\right)\end{equation*}
\State Let $\tilde N_{T^{(i)}} := \{T_j^{(i)} : B_{T^{(i)}}^{j} \ne 0\}$, $\tilde Z_{T^{(i)}} := \{T_j^{(i)} : B_{T^{(i)}}^{j} = 0\}$, where, \newline
\hspace*{5em} $T^{(i)} = \tilde N_{T^{(i)}} \cup \tilde Z_{T^{(i)}}$
\State Define:
\begin{equation*} 
B_{T^{(i)}} := \frac{|\tilde N_{T^{(i)}}|}{|T^{(i)}|}
\end{equation*}
\EndFor
\begin{equation*}
    B_{Avg(G)} = \frac{1}{4} \sum_{i=1}^4 B_{T^{(i)}}
\end{equation*} 
\end{algorithmic}[1]
\caption{Computing triadic balance for a signed directed network}\label{matching1}
\label{meth:algorithm}
\end{algorithm}

\section{Empirical analysis and method}
Real-world communication, verbal or nonverbal, written or visual, involves various types of explicit and implicit relationships, such as like versus dislike, or trust versus distrust. To collect data on communication networks, researchers have used different methods \cite{bernard1990comparing}, such as observations \cite{newcomb_acquaintance_1961}, surveys \cite{sampson1969novitiate}, and text analysis \cite{diesner2015words,culotta2006}.

In this study, we leverage two approaches to label given edges with signs. 
To validate our proposed method for calculating balance in directed triads, we construct communication networks from different social contexts; two business organizations (Enron email dataset, Avocado Research Email collection), and decision-making teams in virtual simulations.
For the two email datasets, we leverage Natural Language Processing (NLP) methods to extract two types of edge signs from text data exchanged between the nodes (authors) that form a dyad: moral values (virtue or vice) and sentiment (positive or negative), as explained further below. 
For the third dataset (decision-making teams), a survey was conducted to extract edge signs with respect to perceived trust between pairs of individuals. 
In the following sections, we explain in detail how we constructed the networks and signed their edges.

\subsection{Network construction and edge labeling Using NLP}
\subsubsection{Data} \label{sec:data1}
The Enron email data is a large-scale, temporal dataset from a global, U.S. based, former energy brokerage that went bankrupt in 2001. The communication (email) dataset of 158 employees was released in 2002 by the FERC \cite{diesner2005communication,investigationpa02}. The original dataset went through various edits and modification over the years. In this study, we use the latest release of the dataset from 2015\footnote{\url{https://www.cs.cmu.edu/~./enron/}}. The Enron dataset is of special importance in the social network community since it provides real-world organizational communication data from over a span of 3.5 years. 

The Avocado Research Email Collection \citet{oard2015avocado} is provided by the Linguistic Data Consortium\footnote{\url{https://catalog.ldc.upenn.edu/LDC2015T03}} and consists of emails between 279 accounts of a defunct information technology company referred to as ``AvocadoIT'', a pseudonym assigned for anonymity. 
The dataset consists of calendars, attachments, contacts, reports and emails. In this study, we focus on the email communication. 

For both datasets, we preprocessed the emails and removed numbers, punctuation, time, and date. In addition, we removed the emails threads tagged as ``Original Email'', and only use the latest communication between the sender and receiver(s). We kept all forwarded emails tagged as ``Forwarded Emails'' for our analysis since we believe that the senders found this information relevant for the receivers. Furthermore, We removed all emails sent by list-serves as well as spam-like email addresses, e.g., ``outlook.team@enron.com'', and ``all@avocadoit.com''. We identified these email addresses by analyzing a random sample of both the Enron and Avocado data sets.

\subsubsection{Edge labeling based on morality and sentiment}
For this study, we label links (emails) based on their valence with respect to moral values and sentiment. This approach is based on the premise that people's language use can reflect their cultural, economic, and ideological backgrounds \cite{triandis1989self}. Language is one of the most powerful means through which people demonstrate their implicit or explicit beliefs, feelings, and moral values \cite{fiske1993controlling}. Differences in people's feelings, opinions, and moral or personal values may be the sources of tension and conflict in relationships and groups. Therefore, extracting and analyzing these relationships from language exchanged between network participants can help to better understand structure and balance in social networks. 

To capture moral values in our email data sets, we leveraged the Moral Foundations Theory (MFT) \cite{graham2013moral, graham2009liberals}. MFT can help to capture people's spontaneous reactions and categorizes human behavior into five basic principles (fairness/cheating, care/harm, authority/subversion, loyalty/betrayal, and purity/degradation) that are characterized by opposing values (virtues and vices). The Moral Foundation Dictionary (MFD) enables the measurement of MFT based on text data by associating 324 words with virtues and vices from the MFT \cite{graham2009liberals,graham2013moral}.
To extract moral values from our email data, we used an enhanced version of MFD\footnote{\url{https://doi.org/10.13012/B2IDB-3805242_V1.1}} as developed, introduced and validated in \cite{rezapour2019enhancing,atillinoisdatabankIDB-3957440}. Compared to the original MFD, the enhanced lexicon consists of about 4,636 terms that were syntactically disambiguated, and manually pruned and verified. In order to analyze balance, we need to extract and label edge signs. For this purpose and study, we only consider the polarity of moral words (virtue or vice), and do not take the moral dimensions into consideration.

The second NLP method we used for labeling links with signs is sentiment analysis; a technique commonly used for understanding people's opinions and affective state \cite{pang2008opinion}. 
The basic task with sentiment analysis is to identify the polarity of communication or discourse, and to label pieces of text data as positive, negative, or neutral. 
To identify the sentiment of each email, we leverage the Subjectivity Lexicon, a widely adopted and previously evaluated sentiment lexicon developed by Wiebe and Riloff \cite{wiebe2005creating}. This lexicon contains a total of 8,222 syntactically disambiguated words that are tagged as having negative, positive, or neutral polarity.

Furthermore, we domain-adjusted both lexicons to align them with the language of our email datasets. For this purpose, we first extracted a list of top (informative) words from the Enron and Avocado datasets (separately). To do that, we used $TF-IDF$ scores (a.k.a. Term Frequency–Inverse Document Frequency), which are high for words that occur many times in a few documents, and reflect the importance of each word in a dataset.
Next, we trained two human annotators to (1) remove overly common words (false positives) from the lexicons, and (2) add relevant but missing words (false negatives) to the lexicons. 
Using the list of top words, the annotators checked if the extracted words did already exist in the lexicons, and whether their prior polarity and part of speech (POS) were appropriate given the context of the email datasets. If a word did not exist in the lexicons, and both coders found it appropriate for the purpose of this study, the word was added to the respective lexicons. If the word was not found suitable and or already exist in a lexicon, we remove the entry from the lexicon. Finally, if a word did exist in a lexicon but both coders agreed on changing the polarity or POS of the word, we modified the entry respectively in the domain adapted lexicon.

After preprocessing the emails (see Section \ref{sec:data1}) from both datsets and domain-adapting the two lexicons, we used $spaCy$ \cite{honnibal2017spacy}, a $Python$ library, to split the emails into sentences, tokenize the sentences into words, and tag each word with its respective POS. 
In order to assign an edge sign that indicate a virtue (+) or vice (-) based on morality of the communication sent from node $P$ to $O$, if a word and its POS match an entry in the Enhanced Morality Lexicon, we counted the number of words for either morality polarity value (+, -), and tagged the sentence with the moral polarity with the highest count. 
Similarly, for edge sign labeling based on sentiment analysis of email content, for any word that matches an entry in our domain adapted Subjectivity Lexicon in surface form and POS, we logged a match, count all matches per sentence and sentiment class (positive, negative, or neutral), and tagged each sentence with the majority class. 
We also checked each sentence for negation using the $NLTK$ package \cite{loper2002nltk}. If a negation was found, we flipped the morality or sentiment polarity to its opposite value; e.g., for morality, from virtue to vice. 
Finally, we aggregate the moral or sentiment polarity of all sentences per email, and normalized the score by the number of sentences per email. 

After tagging morality and sentiment in both email datasets, we constructed four directed edgelists (Avocado Morality, Enron Morality, Avocado Sentiment, and Enron Sentiment) in which email addresses are nodes (senders are source nodes, and receivers are target nodes), emails sent from a node to another node are directed edges, morality or sentiment scores (normalized counts of each email) are the weights of each edge, and morality or sentiment polarity (+, -) are the signs of the edges. If an email does not contain any word that matches a lexicon entry, the email is not considered in the respective edgelists, therefore, an edge can be present in the sentiment edgelists but not in the morality edgelists. 
To construct the networks, we create one edge between each two nodes. Furthermore, if two nodes have more than one edge (email communication), we normalize the morality or sentiment scores of all edges between them.

\subsection{Network construction and edge labeling based on survey data}
\subsubsection{Data}
We leveraged data from an experiment that examined decision-making processes in teams. The experiment involved 18 four-person teams, and each team needed to complete a mission on a virtual simulation platform (Virtual Battlespace 2\footnote{https://bisimulations.com/products/vbs2}). Each four-person team consisted of two smaller units called ``phantom'' and ``stinger''. Each unit had two team members, one commander and one driver. 
The mission entailed navigating a course where teams needed to (a) keep a log of landmarks visited, (b) successfully overcome hazards, and (c) coordinate with the other team in the squad to reach a given rendezvous point before fighting insurgents ahead \cite{pilny2014dynamic}. 
After each mission, team members were asked to rate each other on ``the extent to which you trust your team member within the squad'' on a scale from 1 to 5; with 1 being ``not all all'' and 5 being ``to a very great extent''. 

\subsubsection{Edge labeling based on trust}
The reported trust data from the 18 four-person teams were transformed into a directed network edgelist. An edge was given a positive (+) sign if a team member reported trust towards another member with a rating of 3 or above. Otherwise, an edge was given a negative (-) sign. Note that triads are completely connected within each team (triad type 300) since every team member was asked to report trust scores for all other team members. 

\subsection{Balance analysis based on edge signs and transitivity}
\subsubsection{Edgelist preparation}
One challenge with the Enron dataset is that individuals may have more than one email address \cite{diesner2005communication}. For instance, \textit{``Kenneth Lay''} as the founder, CEO, and chairman of Enron was using the following email addresses: \textit{``kennethlay@enron; klay@enron.com; k.lay@enron.com; k\_lay@enron.com; \\ ken.lay@enron.com; ken\_lay@enron.com; kenlay@enron.com; kenneth.l.lay@\\enron.com; kenneth.lay@enron.com; kenneth\_lay@
enron.com; kenneth\_lay@\\enron.net; kennethlay@enron.com; klay@enron.com; kllay@enron.com; lay.\\kenneth@enron.com; layk@enron.com; ssskenneth.lay@enron.com''}.

After extracting the edge signs, we first converted the email addresses into actual names of the people in the Enron dataset \cite{diesner2005communication,diesner2015little}. In order to disambiguate the email addresses, we leveraged the work by \cite{diesner2005communication}, which includes disambiguated names and email addresses of 558 employees of Enron. 
The final number of nodes and edges in both the Enron sentiment and Enron morality dataset are shown in table \ref{tab:Table2}. The difference in the number of nodes and edges of the two lists is because of the availability of sentiment and morality words in the emails.      

For the Avocado dataset, to be consistent with the Enron dataset, we only considered emails that were sent to or from corporate email addresses (emails ending to \textit{@avocadoit.com}). The number of nodes and edges for the Avocado datasets are shown in table \ref{tab:Table2}.  

For the trust dataset, preprocessing involved converting survey data into directed network edgelists that represent trust scores between every pair of team members. We then recorded the trust score from Likert scale values to binary values; with 1 meaning that ``trust is present'' and 0 being ``trust is not present''. The value for trust score of 1 includes $n \geq 3$, and the value for trust score of 0 includes $n \leq 2$.

\subsubsection{Balance calculation}
After cleaning the edge lists and disambiguating names and email addresses, we used $NetworkX$, a $Python$ library, to remove self-loops, isolates, and pendants. In addition, we removed edges with neutral (0) scores. 

To analyze balance in triads, we follow the steps explained in section \ref{theory} and Algorithm \ref{meth:algorithm}. Moreover, we first extract instances of four transitive triads (030T, 120D, 120U, and 300), and analyze balance within each triad with respect to their semicycles. 
Tables \ref{tab:Table3}, \ref{tab:Table4}, \ref{tab:Table5}, \ref{tab:Table6}, and \ref{tab:Table7} show the final count and ratio of completely balanced, partially balanced, and completely imbalanced transitive triads in each dataset.   

\begin{table}[htb]
\centering
\resizebox{\textwidth}{!}{%
\begin{tabular}{|l|c|c|c|c|c|}
\hline
\multirow{2}{*}{\textbf{Network Measures}} & \multicolumn{2}{c|}{\textbf{Enron}} & \multicolumn{2}{c|}{\textbf{Avocado}} & \textbf{Decision\_Teams} \\ \cline{2-6} 
 & \textbf{Morality} & \textbf{Sentiment} & \textbf{Morality} & \textbf{Sentiment} & \textbf{Trust} \\ \hline
\textbf{\# of nodes} & 494 & 491 & 452 & 402 & 72 \\ \hline
\textbf{\# of edges} & 7520 & 7344 & 22953 & 23519 & 216 \\ \hline
\textbf{Transitivity} & 0.21 & 0.2 & 0.5 & 0.5 & 1 \\ \hline
\textbf{\begin{tabular}[c]{@{}l@{}}Degree \\ Centralization\end{tabular}} & 0.061 & 0.06 & 0.22 & 0.29 & 2 \\ \hline
\textbf{Density} & 0.031 & 0.03 & 0.11 & 0.14 & 1 \\ \hline
\textbf{\begin{tabular}[c]{@{}l@{}}Average Path\\  Length\end{tabular}} & 2.53 & 2.56 & 1.7 & 1.6 & 1 \\ \hline
\textbf{\begin{tabular}[c]{@{}l@{}}Clustering \\ Coefficient\end{tabular}} & 0.46 & 0.46 & 0.62 & 0.68 & 1 \\ \hline
\textbf{\# of Components} & 1 & 1 & 1 & 1 & 1 \\ \hline
\textbf{\begin{tabular}[c]{@{}l@{}}\# of node in largest\\  component\end{tabular}} & 494 & 491 & 452 & 402 & 72 \\ \hline
\end{tabular}%
}
\caption{Descriptive network measures of (1) Enron, (2) Avocado, and (3) Decision-Teams networks}
\label{tab:Table2}
\vspace{-5mm}
\end{table}

\section{Results}
\subsection{Descriptive network measures}
\begin{figure}[htp]
\centering
\includegraphics[width=0.80\textwidth]{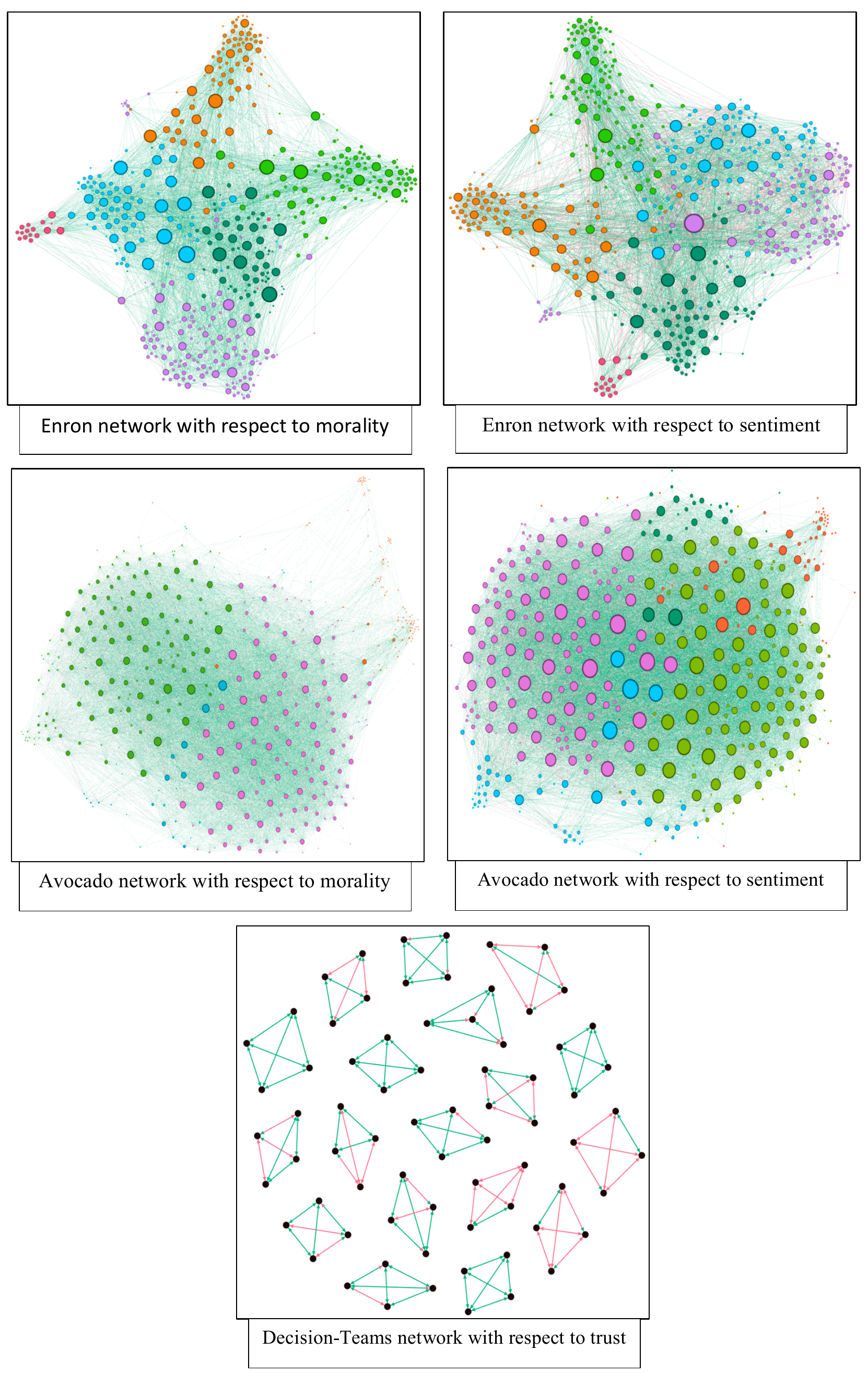}
\caption{Enron, Avocado, and Decision-Teams networks. Clustering method: Louvain modularity. Edge colors: green = positive, red = negative}
\label{fig:Fig3}
\end{figure}

Table \ref{tab:Table2} shows structural characteristics of the three networks, (1) Enron, (2) Avocado, and (3) Decision-Teams. Network visualizations of the networks are also shown in Figure \ref{fig:Fig3}. Enron's networks for morality and sentiment are both sparse, with low amounts of transitive relations among nodes. Low degree centralization in both Enron networks also signifies that there is a limited number of nodes with frequent emailing activity. For the Avocado networks, we observe higher density and transitivity than for Enron. Degree centralization is also higher in Avocado than in Enron, signaling the possible presence of significant number of nodes who are active in sending and receiving emails. Overall, the Avocado networks are denser than the Enron networks, though they have a similar number of nodes. Furthermore, we expect a higher number of triads in Avocado than in Enron, as high density suggests higher occurrences of closed triads. 
For the Decision-Teams network, experimental conditions produce a fully-connected graph. 

\subsection{Balance analysis}
Tables \ref{tab:Table3} and \ref{tab:Table4} present balance results for the Enron networks. The morality network has a overall balance ratio of 92.37\%. All four triad types have high balance ratios, ranging from 91.47\% - 93.89\%. 
The sentiment network has an overall balance ratio of 67.50\%, with triad 300 having the highest balance ratio (69.94\%) and triad 120U having the lowest balance ratio (64.36\%). The prevalence of balanced triad 300s shows that balance is present in situations where individuals initiate and reciprocate email communication. One notable difference in triad 300 counts between morality and sentiment networks is that there is higher partial balance in sentiment network, as opposed to morality network where complete balance is higher. This indicates that while three individuals are fully connected in terms of sending receiving emails, there may be inconsistencies with the sentiment exchanged, but not so much with morality. 

Enron's morality and sentiment networks have similar triadic profiles, in which triad 030T occurs most frequently and is often balanced (91.47\% for morality, 67.46\% for sentiment). In the context of this dataset, the 030T triad represents triples of individuals who are bounded by a certain ``local hierarchy'' - $P$ sends an email to $O$, who then sends an email to $X$, then followed by $P$ sending an email to $X$ as well. Such behavior implies a hierarchy, where both $P$ and $O$ initiate communication with $X$, and $X$ may be at a higher level of influence (consistent with the assumptions of Ranked Clusters model, see \cite{de1999sign}). High counts of balanced triads of type 030T also indicates a strong correlation between transitivity and balance at the triad level of the network. Triad 300 represents complete and reciprocated interaction among three individuals, and these communications are carried out with less tension.
High triad 030T counts also means that there is lower reciprocity at the triad level. This insight has implications for professional email communication and practices for companies in crisis as we observe more instances of initiating emails to other individuals and less reciprocity (i.e., replying) in exchanging emails.  
In addition, we also observe high counts of triads of type 120U, which indicate information reporting (120U, $P$ and $O$ reporting up to $X$), but not of type 120D, which indicate the act of passing down information. This finding suggests hierarchical information flow at Enron, where email communication is initiated by employees and sent to personnel at different levels in the organization. 

\begin{table}[H]
\centering
\resizebox{\textwidth}{!}{%
\begin{tabular}{|c|r|r|r|r|r|r|}
\hline
\begin{tabular}[c]{@{}l@{}}Enron\_\\Morality\end{tabular} & Type & Count & \begin{tabular}[c]{@{}l@{}}Completely\\Balanced\end{tabular} & \begin{tabular}[c]{@{}l@{}}Partially\\ Balanced\end{tabular} & \begin{tabular}[c]{@{}l@{}}Completely\\Imbalanced\end{tabular} & \begin{tabular}[c]{@{}l@{}}Balance \\Ratio ($B_{T^{(i)}}$)\end{tabular} \\ \hline
\multirow{4}{*}{\begin{tabular}[c]{@{}l@{}}Transitive\\Triads\end{tabular}} & 030T & 4514 & 4129 & 0 & 385 & 91.47\% \\ \cline{2-7}
 & 120D & 2390 & 2120	& 161 &109	& 92.07\% \\ \cline{2-7} 
 & 120U & 3615 &	3244 & 167 & 204 & 92.04\% \\ \cline{2-7} 
 & 300 & 3056 & 2696 & 339 & 21 & 93.89\% \\ \hline
\multicolumn{2}{|c|}{Total} & 13575 & 12189 & 667 & 719 & \begin{tabular}[c]{@{}l@{}}$B_{Avg(G)}$ = \\92.37\% \end{tabular}\\ \hline
\end{tabular}
}
\caption{Balance counts with respect to \textbf{morality} in Enron network}
\label{tab:Table3}
\end{table}

\begin{table}[H]
\centering
\resizebox{\textwidth}{!}{%
\begin{tabular}{|c|r|r|r|r|r|r|}
\hline
\begin{tabular}[c]{@{}l@{}}Enron\_\\Sentiment\end{tabular} & Type & Count & \begin{tabular}[c]{@{}l@{}}Completely\\Balanced\end{tabular} & \begin{tabular}[c]{@{}l@{}}Partially\\ Balanced\end{tabular} & \begin{tabular}[c]{@{}l@{}}Completely\\Imbalanced\end{tabular} & \begin{tabular}[c]{@{}l@{}}Balance \\Ratio ($B_{T^{(i)}}$)\end{tabular} \\ \hline
\multirow{4}{*}{\begin{tabular}[c]{@{}l@{}}Transitive\\Triads\end{tabular}} & 030T & 4238 & 2859 & 0 & 1379 & 67.46\% \\ \cline{2-7} 
 & 120D & 2384 & 1333 & 588 & 463 & 68.24\% \\ \cline{2-7}
 & 120U & 3513 & 1775 & 972 & 766 & 64.36\% \\ \cline{2-7}
 & 300 & 3056 & 1312 & 1605 & 139 & 69.94\% \\ \hline
\multicolumn{2}{|c|}{Total} & 13191 & 7279 & 3165 & 2747 & \begin{tabular}[c]{@{}l@{}}$B_{Avg(G)}$ = \\67.50\% \end{tabular}\\ \hline
\end{tabular}
}
\caption{Balance counts with respect to \textbf{sentiment} in Enron network}
\label{tab:Table4}
\end{table}

Tables \ref{tab:Table5} and \ref{tab:Table6} show balance results for the Avocado networks. The overall balance ratio for morality is 86.70\%, with triad 030T having the lowest balance ratio (80.74\%), while triad 300 has the highest balance ratio (93.47\%). The overall balance ratio for sentiment is 82.47\%, with the same profile of triad 030T having the lowest balance ratio (76.22\%), and triad 300 having the highest balance ratio (90.28\%). In addition, triad 300 is the most frequently-occurring one in both Avocado networks.  

Similar to the Enron networks, the Avocado networks contains substantially more counts of 120U than 120D. Recurring prominence of 120U triads in email communication networks may indicate the prevalence of information reporting. 
We observe more consistency in balance ratios of the Avocado networks compared to Enron, where balance ratios is only 4.23\% for Avocado, and 24.87\% for Enron. 
For example, six emails exchanged among three managers (triad type 300) all highlighted the virtue of authority (in morality), but one of the emails contained negative sentiment, which influenced the overall balance ratio of sentiment for that particular triad. 
One reason for such inconsistencies in just the Enron networks could be because this company underwent a series of crises which resulted in bankruptcy, and this may have had profound effects on the sentiment of the emails. 

The overall balance ratio in Avocado's morality network (86.70\%) is slightly lower than Enron's morality network (92.37\%), possibly because Avocado's network size is three times larger, hence providing more opportunities to develop balance (or in this case, imbalance) among triads. On the other hand, Avocado's sentiment network has higher balance ratio (82.47\%) than Enron's sentiment network (67.50\%), indicating that there may be less tension in the emails exchanged between Avocado employees compared to Enron.    
Another difference between Avocado and Enron s is that Avocado networks contain higher proportions of 300s triads (72\% for morality; 74\% for sentiment). In contrast to Enron networks, which contain mostly 030T triads, Avocado networks are more tightly-connected with frequent and reciprocated communications. 
With respect to triad counts, Enron's morality and sentiment networks have a similar total number of triads (13,575 and 13,191, respectively). Avocado's morality network has notably less triads than the sentiment network (174,434 and 196,333, respectively). This difference in triad counts indicates that individuals at Avocado use more sentiment terms in their email changes. 

\begin{table}[H]
\centering
\resizebox{\textwidth}{!}{%
\begin{tabular}{|c|r|r|r|r|r|r|}
\hline
\begin{tabular}[c]{@{}l@{}}Avocado\_\\ Morality\end{tabular} & Type & Count & \begin{tabular}[c]{@{}l@{}}Completely\\Balanced\end{tabular} & \begin{tabular}[c]{@{}l@{}}Partially\\ Balanced\end{tabular} & \begin{tabular}[c]{@{}l@{}}Completely\\Imbalanced\end{tabular} & \begin{tabular}[c]{@{}l@{}}Balance \\Ratio ($B_{T^{(i)}}$)\end{tabular} \\ \hline
\multirow{4}{*}{\begin{tabular}[c]{@{}l@{}}Transitive\\Triads\end{tabular}} & 030T & 8787 & 7095 & 0 & 1692 & 80.74\% \\ \cline{2-7}
 & 120D & 14111 & 11627 & 882 & 1602 & 85.52\% \\ \cline{2-7}
 & 120U & 26165 & 22257 & 1047 & 2861 & 87.06\% \\ \cline{2-7}
 & 300 & 124371 & 109528 & 13203 & 1640 & 93.47\% \\ \hline
\multicolumn{2}{|c|}{Total} & 173434 & 150507 & 15132 & 7795 & \begin{tabular}[c]{@{}l@{}}$B_{Avg(G)}$ = \\86.70\% \end{tabular}\\ \hline
\end{tabular}
}
\caption{Balance counts with respect to \textbf{morality} in Avocado network}
\label{tab:Table5}
\end{table}

\begin{table}[H]
\centering
\resizebox{\textwidth}{!}{%
\begin{tabular}{|c|r|r|r|r|r|r|}
\hline
\begin{tabular}[c]{@{}l@{}}Avocado\_\\ Sentiment\end{tabular} & Type & Count & \begin{tabular}[c]{@{}l@{}}Completely\\Balanced\end{tabular} & \begin{tabular}[c]{@{}l@{}}Partially\\ Balanced\end{tabular} & \begin{tabular}[c]{@{}l@{}}Completely\\Imbalanced\end{tabular} & \begin{tabular}[c]{@{}l@{}}Balance \\Ratio ($B_{T^{(i)}}$)\end{tabular} \\ \hline
\multirow{4}{*}{\begin{tabular}[c]{@{}l@{}}Transitive\\Triads\end{tabular}} & 030T & 8577 & 6538 & 0 & 2039 & 76.22\% \\ \cline{2-7}
 & 120D & 14276 & 10816 & 1408 & 2052 & 80.69\% \\ \cline{2-7}
 & 120U & 28615 & 22802 & 1725 & 4088 & 82.69\% \\ \cline{2-7}
 & 300 & 144865 & 118673 & 23870 & 2322 & 90.28\% \\ \hline
\multicolumn{2}{|c|}{Total} &196333 & 158829 & 27003 & 10501 & 
\begin{tabular}[c]{@{}l@{}}$B_{Avg(G)}$ = \\82.47\% \end{tabular}\\ \hline
\end{tabular}
}
\caption{Balance counts with respect to \textbf{sentiment} in Avocado network}
\label{tab:Table6}
\end{table}

\subsubsection{Decision-teams}
Table \ref{tab:Table7} shows balance counts of the 18 decision-making teams. Given the experimental condition, all teams' networks are completely connected, resulting in 432 semicycles, all of which are embedded within 72 triads of type 300s. The overall balance in this network is 72.69\%. Specifically, 60\% of triads (43 out of 72) are partially balanced, 40\% (29 out of 72) are completely balanced, and 1\% (1 out of 72) is completely imbalanced. This shows that many triads contain some amount of tension, but not significant enough that they become completely imbalanced.  

\begin{table}[htb]
\centering
\resizebox{\textwidth}{!}{%
\begin{tabular}{|c|r|r|r|r|r|r|r|}
\hline
\begin{tabular}[c]{@{}l@{}}Trust \end{tabular} & Type & Count & \begin{tabular}[c]{@{}l@{}}Completely\\Balanced\end{tabular} & \begin{tabular}[c]{@{}l@{}}Partially\\ Balanced\end{tabular} & \begin{tabular}[c]{@{}l@{}}Completely\\Imbalanced\end{tabular} & \begin{tabular}[c]{@{}l@{}}Balance \\Ratio ($B_{T^{(i)}}$)\end{tabular} \\ \hline
\multirow{4}{*}{\begin{tabular}[c]{@{}l@{}}Transitive\\Triads\end{tabular}} & 030T & 0 & 0 & 0 & 0 & 0\% \\ \cline{2-7}
 & 120D & 0 & 0 & 0 & 0 & 0\% \\ \cline{2-7}
 & 120U & 0 & 0 & 0 & 0 & 0\% \\ \cline{2-7}
 & 300 & 72 & 29 & 43 & 1 & 72.69\% \\ \hline
\multicolumn{2}{|c|}{Total} & 72 & 29 & 43 & 1 & \begin{tabular}[c]{@{}l@{}}$B_{Avg(G)}$ = \\ 72.69\% \end{tabular}\\ \hline
\end{tabular}
}
\caption{Balance counts with respect to \textbf{trust} in Decision-Teams network}
\label{tab:Table7}
\vspace{-3mm}
\end{table}

\subsection{Sign analysis of semicycles}
All three networks contain higher proportions of positive than negative edges. Equivalently, higher proportions of positive semicycles are observed. The results for signed triads (\ref{tab:Table8} for Enron, table \ref{tab:Table9} for Avocado, and table \ref{tab:Table10} for trust) show the higher instances of positive ties within semicycles, which explains higher occurrences of both $+++$ and $++-$ semicycles than semicycles that contain higher counts of negative ties. Our findings are consistent with prior work by Doreian and Krackhardt \cite{doreian2001pre} and Davis \cite{davis1979davis}, who all found transitive to be a pre-condition for balance when both $P$ $\rightarrow$ $O$ and $O$ $\rightarrow$ $X$ are positive. Lescovec, Huttenlocher, and Kleinberg \cite{leskovec2010signed} also empirically observed a majority of all-positive semicycles in three real-world social networks; with the proportion of positive semicycles ranging from 70\% to 87\%. 

The differences in sign counts for morality and sentiment are more salient in the Enron networks (table \ref{tab:Table8}) than in the Avocado networks (table \ref{tab:Table9}). The proportions of $+++$ and $++-$ semicycles are similar in the Enron sentiment network, indicating a higher amount of imbalance in this network. The $++-$ semicycle represents a unique type of tension that we frequently observed in the Enron data; e.g., when \textit{``Jeff Skilling''} sent a positive email to \textit{``Rebecca Mark''} (Head of Enron International), \textit{``Rebecca''} sent a positive email to \textit{``Kenneth Lay''} (CEO and Chairman of Enron), but \textit{``Kenneth Lay''} in turn sent a negative email to \textit{``Jeff Skilling''}. This case exemplifies a violation of transitivity and structural balance, such that the link between \textit{``Kenneth Lay''} and \textit{``Jeff Skilling''} is a source of tension between wihtin a triad. Another interpretation could be that organizational emailing etiquette is generally more positive, with the occasional presence of negative emails within dyads. In fact, all-negative semicycles are rare (about 0.5\% in Enron networks, and 0.03\% in Avocado networks), suggesting that it is not common to engage in chains of negative emails.

\begin{table}[htb]
\centering
\begin{tabular}{|c|r|r|r|r|}
\hline
Enron & \multicolumn{2}{c|}{Morality} & \multicolumn{2}{c|}{Sentiment} \\ \hline
Semicycle type & Counts & Ratio-Total & Counts & Ratio-Total \\ \hline
+ + + & 32202 & 0.92 & 19830 & 0.58 \\ \hline
+ + - & 2450 & 0.07 & 10419 & 0.30 \\ \hline
+ - - & 199 & 0.006 & 3630 & 0.11 \\ \hline
- - - & 9 & 0.0003 & 489 & 0.01 \\ \hline
Total & 34860 & 1 & 34368 & 1 \\ \hline
\end{tabular}
\caption{Types of signed semicycles for Enron morality and sentiment}
\label{tab:Table8}
\end{table}

\begin{table}[htb]
\centering
\begin{tabular}{|c|r|r|r|r|}
\hline
Avocado & \multicolumn{2}{r|}{Morality} & \multicolumn{2}{r|}{Sentiment} \\ \hline
Semicycle type & Counts & Ratio-Total & Counts & Ratio-Total \\ \hline
+ + + & 768926 & 0.92 & 851297 & 0.89 \\ \hline
+ + - & 61069 & 0.07 & 92938 & 0.10 \\ \hline
+ - - & 5419 & 0.006 & 10340 & 0.01 \\ \hline
- - - & 151 & 0.0002 & 409 & 0.0004 \\ \hline
Total & 905047 & 1 & 954984 & 1 \\ \hline
\end{tabular}
\caption{Types of signed semicycles for Avocado morality and sentiment}
\label{tab:Table9}
\end{table}
The sign counts for the decision teams networks (see table \ref{tab:Table10}) are distinct from the Avocado and Enron email communication networks. Similar to the Enron and Avocado email networks, decision-teams trust network also has the highest proportion of $+++$ semicycle. The primary difference with this particular network is that $+--$ semicycles are more prevalent than $++-$ semicycles. The prevalence of two balanced semicycle types in this network is evidence for the tendency to maintain balance; team members may orient their perceptions of trust towards other team members in ways that potentially reduce tensions within their immediate teams. Specifically with the experimental setup where individuals are split into two teams, we observe a number of cases where team member $P$ of team ``phantom'' trusts (+) member $O$ of the same team, but member $O$ does not trust (-) member $X$ of the ``stinger team'', therefore team member $P$ does not trust (-) member $X$, maintaining a strong sense of trust within the team and low trust outside of the team. This finding is consistent with previous literature on trust in organizations that has shown how trust rather emerges within teams than across teams \cite{ashleigh2001trust,de2010does} as members who actively and frequently work together develop a higher sense of team identity \cite{hogg2012social}.

\begin{table}[t]
\centering
\begin{tabular}{|c|r|r|}
\hline
Semicycle type & Counts & Ratio-Total \\ \hline
+ + + & 196 & 0.45 \\ \hline
+ + - & 91 & 0.21 \\ \hline
+ - - & 118 & 0.27 \\ \hline
- - - & 27 & 0.06 \\ \hline
Total & 432 & 1 \\ \hline
\end{tabular}
\caption{Types of signed semicycles for Decision-Teams trust}
\label{tab:Table10}
\end{table}

\section{Discussion and conclusion}
\label{S:2}
In this paper, we developed a theoretical framework for calculating balance in signed, transitive digraphs, which was essential to appropriately model and study balance in real-world communication networks and other networks where links might be asymmetric. We operationalized and implemented this framework, and applied it to three social networks, namely email communication within an energy firm (Enron network) and an IT company (Avocado), and perceived trust among team members engaged in decision-making tasks (Decision-teams network). Our rationale for testing our approach on different networks was to determine whether mechanisms of structural balance and transitivity hold true across diverse social contexts. Moreover, prior research has mainly examined structural balance in signed and undirected graphs. Our study provided an actionable solution to measure structural balance in signed digraphs, using principles of transitivity to evaluate the directionality between edges. 

Overall, our findings showed that the amount to which a network was balanced was strongly impacted by choices of measuring social relations. When direction of edges was taken into account, along with sign consistency, we expected that the overall balance ratio may be different than findings where only sign consistency was considered \cite{diesner2015little,leskovec2010signed}. Choices of edge type may also have an effect on the overall balance. Our findings showed that each edge type captured a different characteristic of a network, as reflected in the different balance ratios across morality, sentiment, and trust dimensions. While balance ratios for all three edge types were about 70\% and above (balance higher than imbalance), we found that networks labeled with morality as the edge type had the highest balance ratios, while networks labeled with either sentiment or trust as the edge types were notably lower. 

The patterns of structural balance that we discovered across the three networks offer implications for existing communication and organizational networks literature. First, we found that email communication is highly positive in both morality and sentiment. In addition, communication flow was upwards through a hierarchy in the form of information reporting behavior. One implication of this finding was that the observed communication patterns can provide insights into an organization's formal hierarchy, and shed light on the types of influences (e.g., organizational status) that exist to maintain balance in the network. A methodological implication of our findings was that preprocessing text data for network construction impacts balance assessment results. For the sentiment results specifically, overall balance ratios decreased after negation handling and domain adaptation of the applied lexicon. Thus, balance measures may also depend on the researcher's choices about network data preprocessing. This work further expanded research on the impact of human choices about extracting relational data from text data \cite{diesner2015words,diesner2010extraktion}. 

Second, we observed that choices about constructing and aggregating social network data may impact balance ratios. For Enron and Avocado email communication networks, we made an informed choice to normalize all communications between two people (Tables \ref{tab:Table3}, \ref{tab:Table4}, \ref{tab:Table5}, and \ref{tab:Table6}). We performed additional analyses on email datasets and found that choosing the first instance of email communication between two people result in different balance ratios (77.3\% for Avocado-morality, 73.5\% for Avocado-sentiment, 86.7\% for Enron-morality, 61.2\% for Enron-sentiment) compared to considering the last instance of email communication between the same people (76.7\% for Avocado-morality, 64.6\% for Avocado-sentiment, 86.7\% for Enron-morality, 60.0\% for Enron-sentiment). For the decision-teams network data, we also conducted additional balance analysis with a practice mission that preceded the official mission, and found that balance ratio was 58.88\%. These results and considerations highlighted the recurrent problem of constructing static networks from temporal network data, where researchers must make decisions on either aggregating or disregarding instances. These solutions may result in biasing the overall balance ratio of a network. To address this issue, incorporating temporal data (if applicable) into balance analysis will ensure a more comprehensive analysis of networks since it would enable an examination of how networks gravitate towards balance throughout time \cite{diesner2015little,uddin2013dyad}. 

\bibliographystyle{plain}
\bibliography{references}

\end{document}